\documentstyle[pra,aps,twocolumn,epsfig]{revtex}
\draft

\begin{document}                

\def\be{\begin{equation}}
\def\ee{\end{equation}}
\def\ba{\begin{eqnarray}}
\def\ea{\end{eqnarray}}
\def\ban{\begin{eqnarray*}}
\def\ean{\end{eqnarray*}}


\title{On phases in weakly interacting finite Bose systems}
\author{D. J. Dean\cite{dd} and T. Papenbrock\cite{tp}}
\address{Physics Division, Oak Ridge National Laboratory, Oak Ridge,
TN 37831, USA}
\maketitle
\begin{abstract}
We study precursors of thermal phase transitions in finite systems of
interacting Bose gases. For weakly repulsive interactions there is a
phase transition to the one-vortex state. The distribution of zeros
of the partition function indicates that this transition is first
order, and the precursors of the phase transition are already
displayed in systems of a few dozen bosons. Systems of this size do
not exhibit new phases as more vortices are added to the system.
\end{abstract}
\section{Introduction}
The study of vortices in dilute atomic Bose--Einstein condensates has
received a lot of attention in recent years. The condensate wave
function of a vortex state exhibits a quantized circulation of its
velocity field. This state may experimentally be formed, e.g., by
``phase imprinting'' techniques~\cite{JILA} or by directly transferring
angular momentum to the condensating system~\cite{Dalibard}. Current
experiments are performed in the Thomas--Fermi regime of short
coherence length. For a review of recent results see,
e.g., \cite{Fetter}.  The regime of long coherence length is of
interest as well.  Indeed, harmonically trapped Bose systems with
perturbatively weak repulsive interactions display a rich structure
\cite{Rokhsar}. At low ratios of angular momentum $L$ to particle
number $N$, i.e. $L/N \ll 1$, the ground states are dominated by
quadrupole and octupole excitations \cite{Mottelson,Kavoulakis}. The
ground state structure changes smoothly until the one-vortex state
$L/N=1$ is approached. In particular, the ground state wave function
is known analytically and the ground state energy depends linearly on
$L$ for $L/N < 1$ \cite{BP1}.  The one-vortex state is a
Bose--Einstein condensed state where a macroscopic number of particles
carries one quantum of angular momentum each \cite{Wilkin}. The
structure of ground states changes considerably as further angular
momentum is put into the system.  Results from mean-field calculations
exhibit changing symmetries of the developing vortex array which
correspond to changes in the occupation numbers of single-particle
states \cite{Kavoulakis}. This goes along with an increasing
complexity of higher vortex states. The two-vortex state $L/N=2$, for
instance, already  exhibits a complicated structure and is dominated by
excitations that carry 0, 2, and 4 units of angular momentum
\cite{Kavoulakis}. Naturally, the question arises whether the observed
changes in the condensate wave function structure and the formation of
quantized vortices are associated with thermal phase transitions.

For the transition to the one-vortex state, the answer is
affirmative. Wilkin {\it et al.} \cite{Wilkin} showed that the
one-particle reduced density matrix of the one-vortex state has one
eigenvalue of order $N$ and thus meets a criterion for Bose--Einstein
condensation in finite systems; see e.g. \cite{Leggett}. The order of
this phase transition, however, is not known until now. It is the main
purpose of the present paper to classify this phase transition from
its precursors in finite systems.

We recall that small systems do not exhibit phase
transitions. Nevertheless, finite systems may display precursors of
phase transitions. Recently, Borrmann {\it et al.}  proposed that the
distributions of zeros of the canonical partition function ``reveal
the thermodynamic secrets of small systems in a distinct manner''
\cite{Borrmann}. We will apply this method to the case of weakly
interacting Bose systems under rotation.

\section{System and Method}
We consider a system of $N$ bosons confined in a three-dimensional
harmonic trap at total angular momentum $L$ and restrict ourselves to
the sector of maximal magnetic quantum number, i.e. the particles are
in the ground state with respect to excitations along the axis of
rotation. The non-interacting system is highly degenerate. In what
follows, we assume that the repulsive interaction between the bosons is
perturbatively weak and simply lifts this degeneracy.  This yields
sets of now quasi-degenerate states that are separated by multiples
of the oscillator spacing $\hbar\omega$. Under these conditions, the
level spacing between quasi-degenerate states is much smaller than
the oscillator spacing. We are interested in the thermal properties of
the system while keeping the angular momentum $L$ fixed. For
temperatures that are smaller than the oscillator spacing, we may
restrict ourselves to the lowest-lying set of quasi-degenerate states
with approximate energies $E\approx L\hbar\omega$. (We set the ground
state energy of the non-rotating system to zero.) Current experiments
do not work within this low-temperature regime. We recall that the
onset of Bose--Einstein condensation is already observed for
temperatures $kT\sim N^{1/3}\hbar\omega$ \cite{Stringari}. Below we
find that quantitative results concerning thermal phase transitions
are already determined by a few hundred low-lying levels. It thus
seems that one can lift the requirement of perturbatively small
temperatures without facing the inclusion of highly excited states for
the problem under consideration. In the low-temperature regime, the
Hamiltonian of the $N$-boson system with contact interaction reads
\cite{BP1}
\be
\label{ham}
\hat{H}=v_0\sum_{i,j,k,l}
\frac{(k+l)!\,\delta_{k+l}^{i+j}}
{2^{k+l}\left(i!\,j!\,k!\,l!\right)^{1\over 2}}
\,\hat{a}_i^{\dagger} \hat{a}_j^{\dagger} \hat{a}_k \hat{a}_l.
\ee
Here, $v_0$ denotes the strength of the contact interaction. The
operators $\hat{a}_j^{\dagger}$ and $\hat{a}_j$ create and
annihilate one boson in the single-particle state with angular
momentum $j$ with $j=0,1,2\ldots$, respectively.  The operators
\be
\hat{N}=\sum_j \hat{n}_j
\ee
and
\be
\hat{L}=\sum_j j \hat{n}_j
\ee
count the number of particles and the quanta of angular momentum and
have quantum numbers $N$ and $L$, respectively. We have used the
number operators $\hat{n}_j=\hat{a}_j^{\dagger} \hat{a}_j$. Basis
states are denoted as $|n_0,n_1,n_2,\ldots\rangle$, where $n_j$
denotes the number of particles with angular momentum $j$.  Hilbert
space is partition space, and the number of basis states is given by
the number of partitions of $L$ into at most $N$ integers. The
dimension of Hilbert space grows exponentially with increasing $L$ while
being only mildly dependent on $N$ for $L>N$.

Let us now turn to a description of the method proposed in
Ref.\cite{Borrmann,borr2}. It is based on an analysis that Grossmann
and Rosenhauer made for macroscopic systems about three decades ago
\cite{Grossmann}. The canonical partition function $Z({\cal B})$ is
evaluated at complex arguments ${\cal B}=\beta+i\tau$. Since
$Z(\beta)$ is real, it suffices to consider the partition function
$Z({\cal B})$ for arguments in the upper complex plane. Two different
phases of macroscopic systems are separated by a line of zeros which
intersects the real axis at the critical temperature. Further
information about the order of the phase transition is encoded in the
slope of the line at the intersection point and the density of zeros
close to the intersection point. This is physically plausible. We
recall that thermodynamic quantities are given by logarithmic
derivatives of the partition function and thus diverge at its zeros.

In finite systems, the line of zeros reduces to more or less closely
spaced zeros that line up on a curve. Following Ref.~\cite{Borrmann},
one then studies the behavior of the zeros with smallest imaginary
part and of the underlying curve while increasing the number of
particles. This allows one to predict the critical temperature and
order of the phase transition in the infinite system.  Furthermore,
the shape of the curve or the presence of several such curves allows
one to identify different ``phases'' already in finite systems. These
techniques have been used to study precursors of phase transitions in
Bose--Einstein condensates of ideal gases and atomic
cluster~\cite{Borrmann,borr2}. Further applications include the
classification of a pairing phase transition in finite Fermi
systems~\cite{bdhj}.

We compute the canonical partition function from the eigenvalues of
the Hamiltonian~(\ref{ham}). This requires the complete
diagonalization of $\hat{H}$ and is practicable only for moderately
large values of $L$. Zeros of the partition function are determined
conveniently by locating the poles of the specific heat, which
is given by
\begin{equation}
C_v({\cal B})=\frac{\partial^2 \ln Z({\cal B})}{\partial {\cal B}^2}\;.
\end{equation}
For larger
system sizes (i.e. larger values of angular momentum $L$), we restrict
the computation to the lowest hundreds of levels and construct an
approximate partition function only. Nevertheless, we find
sufficiently well-converged zeros of the partition function close to
the real axis of complex temperature ${\cal B}=\beta+i\tau$. This
allows us to present quantitative data concerning the order of the
phase transition in the macroscopic system.

\section{Results}
\begin{figure}
\begin{center}
    \includegraphics*[scale=0.67]{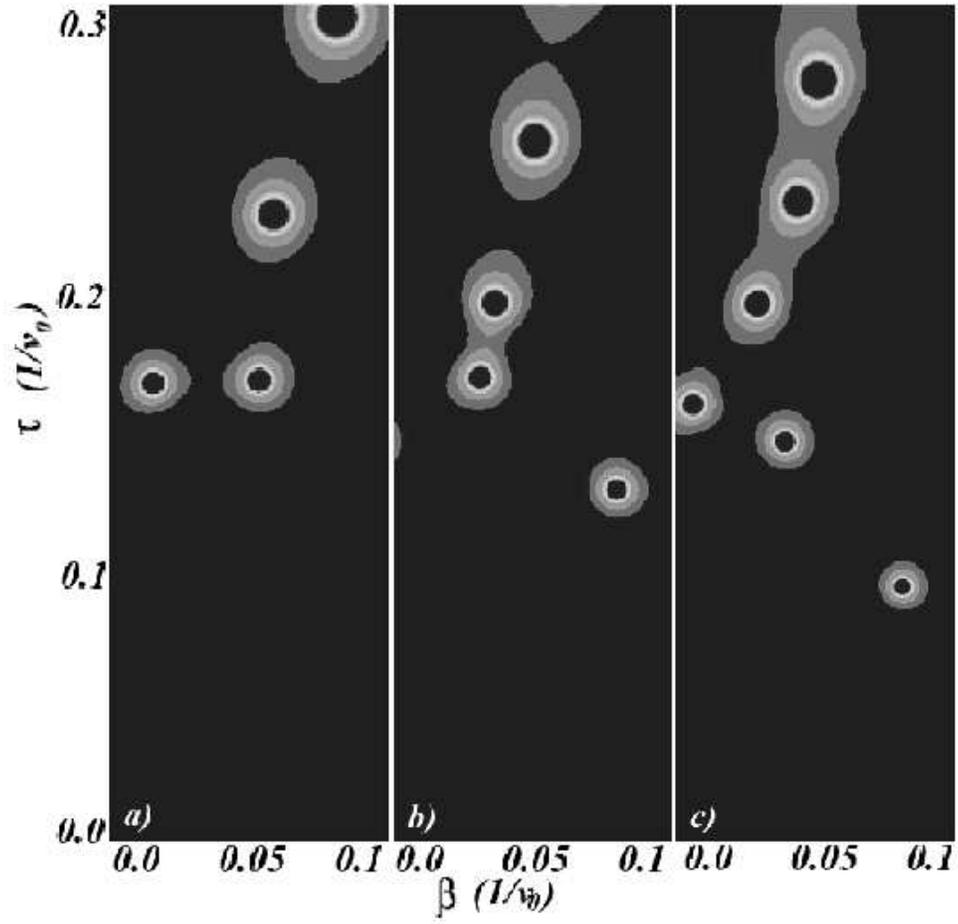}
\end{center}
\caption{Contour plots of the specific heat in the complex temperature plane
for the $N=30$ system at
a) $L=20$, b) $L=25$, and c) $L=30$ units of angular momentum.
The spots indicate the locations of the
zeros of the canonical partition function.}
\label{fig1}
\end{figure}

We investigate the transition to the one-vortex state first. To this
purpose, we fix the number of particles to $N=30$ and fully diagonalize
the Hamiltonian for several values of angular momentum in the range
$0.8 < L/N < 1.07$ around the one-vortex state $L/N=1$. The partition
function is computed from the obtained energy levels. Figure~\ref{fig1}
shows the specific heat as a function of complex ${\cal
B}=\beta+i\tau$. We restricted the plot to include only those zeros
with smallest positive imaginary part. Figure~\ref{fig1} demonstrates
that the zeros approach the real axis with increasing $L$. The closest
encounter is found for the one-vortex state $L=N=30$. This is a
precursor of the condensation into the one-vortex state in the
infinite system and supports earlier
results~\cite{Wilkin,BP1,Kavoulakis}. We are particularly
interested in the order of this phase transition. To this purpose, we
consider the one-vortex state $L=N$ and compute eigenvalues of the
Hamiltonian~(\ref{ham}) for increasing values of particle number
$N$. A complete diagonalization is prohibitively expensive for $N$
exceeding values of about 35. Instead, we restrict ourselves to the
computation of the lowest-lying eigenvalues. These are used for an
approximate construction of the partition function. We found
numerically that its zeros with smallest positive imaginary parts are
already sufficiently well converged when only a few hundred eigenvalues
are included in the computation. We considered systems up to $L=N=55$,
corresponding to a dimension of Hilbert space of the order $4.5\times
10^5$. The relevant eigenvalues are computed numerically using the
ARPACK and PARPACK routines \cite{ARPACK}.

\begin{figure}
\begin{center}
    \includegraphics*[scale=0.67]{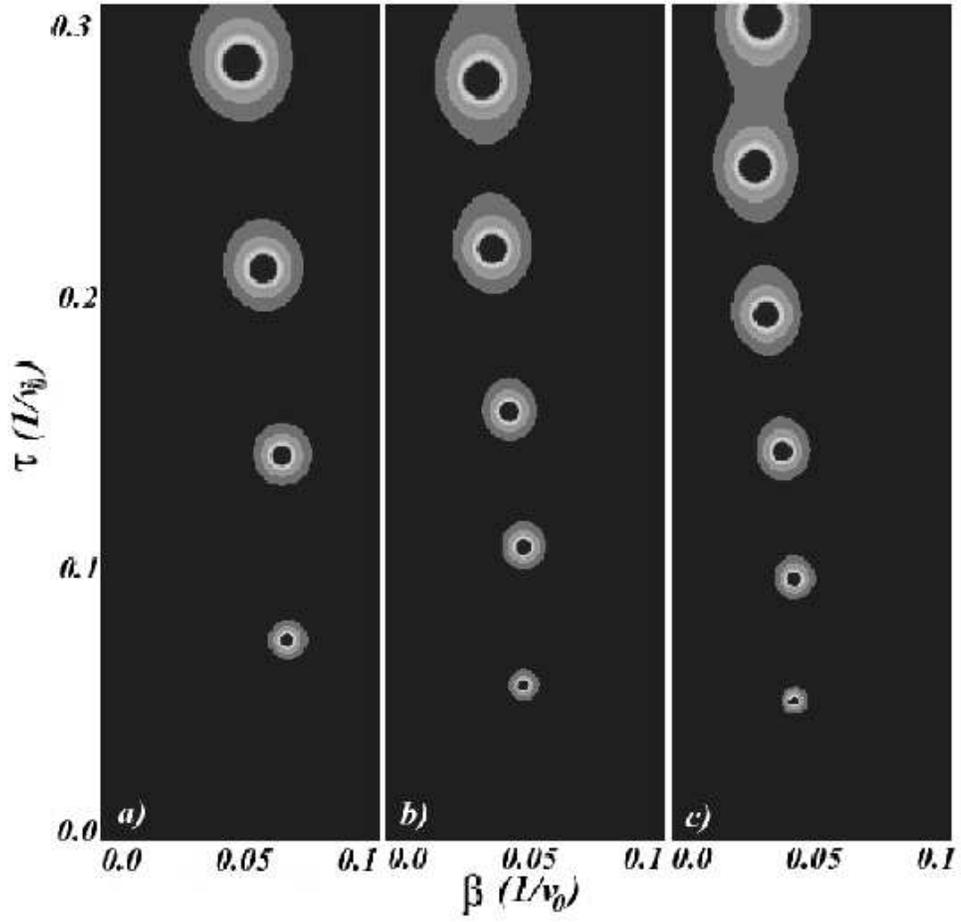}
\end{center}
\caption{Contour plots of the specific heat in the complex temperature plane
for the $L/N=1$ systems with increasing numbers of particles:
a) $N=40$, b) $N=50$, and c) $N=55$.
The spots indicate the locations of the
zeros of the canonical partition function.}
\label{fig2}
\end{figure}

Figure~\ref{fig2} shows the distribution of poles in the complex plane
of the specific heat at the
one-vortex state for different system sizes. These
plots are generated from the lowest-lying 300 eigenvalues. (Increasing
the number of eigenvalues from 300 to 380 yields less than 1\% change in
the numerical results. Thus, the data is sufficiently well converged
for our purposes.) It is clearly seen that the zeros line up and
approach the real axis with increasing system size. The order of the
phase transition is determined as follows \cite{Borrmann}.
The distribution of zeros close to the real axis is approximately described
by three parameters. Two of these parameters reflect the order of the
phase transition, while the third indicates the size of the system.
Let us assume that the zeros lie on a line. We label the zeros according
to their closeness to the real axis. Thus $\tau_1$ reflects the discreteness
of the system. The density of zeros for a given $\tau_k$ is given by
\begin{equation}
\phi\left(\tau_k\right)=\frac{1}{2}
\left(\frac{1}{\mid {\cal B}_k -{\cal B}_{k-1}\mid} +
\frac{1}{\mid {\cal B}_{k+1}-{\cal B}_k \mid}\right)\;,
\end{equation}
with $k=2,3,4,\cdots$ A simple power law describes the density of
zeros for small $\tau$, namely $\phi(\tau)\sim \tau^\alpha$. If we use
only the first three zeros, then $\alpha$ is given by
\begin{equation}
\alpha=\frac{\ln \phi(\tau_3) - \ln\phi(\tau_2)}{\ln\tau_3 -\ln\tau_2}\;.
\end{equation}
The final parameter that describes the distribution of zeros is given
by $\gamma=\tan\nu\sim (\beta_2-\beta_1)/(\tau_2-\tau_1)$.

In the thermodynamic limit, $\tau_1\rightarrow 0$, in which case the
parameters $\alpha$ and $\gamma$ coincide with the infinite system
limits discussed by Grossmann and Rosenhauer \cite{Grossmann}. For
infinite systems, $\alpha=0$ and $\gamma=0$ indicates a first-order
phase transition, while $0<\alpha<1$ and $\gamma=0$ or $\gamma \ne 0$
indicates a second-order transition.  For systems approaching infinite
particle number, $\alpha$ cannot be smaller than zero since this
causes a divergence of the internal energy. In small systems, with
finite $\tau_1$, $\alpha<0$ is possible and is also indicative of a
first-order transition.  We show our results for
$\alpha,\gamma,\tau_1$ in Table \ref{table:params} for the
$N=40,50,52,55$ systems. $\tau_1$ decreases with increasing system
size as a power law. The fit is given by $\tau_{1}=1.15 N^{-1.53}$.
 From the table we note that $\gamma$ is nearly zero, and $\alpha$ is a
small negative number for each system we studied here. The critical
temperature $kT_c$ is approximately given by $1/\beta_1$. Our data
suggests that $kT_c \sim v_0 N^{1.1\ldots 1.4}$. The $N$-dependence
differs considerably from what is found for the ideal Bose gas in
three-dimensional traps \cite{Stringari}.

We thus find that the phase transition to the one-vortex state is
first order. This is the main result of this work and combines well
with previous results found for the energy and wave function
structure. We recall that the ground state energy of the $N$-boson
system exhibits a kink at $L=N$, and that the wave function structure
changes strongly when increasing $L$ beyond $N$ \cite{BP1,Kavoulakis}.

Are there further indications of phase transitions as one approaches
the two-vortex state?  Mean-field results by Kavoulakis {\it et
al.}~\cite{Kavoulakis} show that the structure of ground states
changes considerably at certain ratios of $L/N$. At $L/N\approx 1.75$
the mean-field density develops a two-fold symmetry, and the
expectation $n_j$ for occupation of single-particle orbitals with
odd $j$ vanishes at $L/N \approx 1.75$.  A second change in ground
state structure occurs at $L/N\approx 2.03$ where the two-fold
symmetry of the mean-field density changes into a three-fold
symmetry. This is accompanied by a macroscopic occupation of
single-particle orbitals with angular momentum $j=0, 3, 6, 9$. We
numerically computed the ground states of Hamiltonian~(\ref{ham}) for
systems with moderate particle numbers and confirm these results. We
also computed the partition function from the fully diagonalized
Hamiltonian for $L=30$ and $N$ ranging from 10 to 20.  However, the
distribution of zeros of the partition function does not indicate a
thermal phase transition. In particular, we do not find zeros that
start to line up and approach the real axis at the corresponding
values of $L/N$. This is most likely due to the small system size.

Let us finally turn to Bose--Einstein condensates with attractive
interaction.  In this case, the ground state is the (Bose--Einstein
condensed) state of the non-rotating system which executes pure
center-of-mass motion, i.e. all angular momentum is carried by the
motion of the center of mass \cite{Wilkin,Mottelson}. A direct
calculation of the one-particle reduced density matrix indicates the
presence of a condensate that is fragmented over several states
\cite{Wilkin}. Only when the center-of-mass motion is separated does one
find the condensation into one state \cite{Pethick}. The distribution
of zeros of the partition function does not indicate a phase
transition. This remains the case also when considering systems of size
$L,N\approx 50$.

In summary, we have investigated phases in weakly interacting
Bose--Einstein condensates. The phase transition to the one-vortex
state is first order, and its precursors are clearly seen in the
distribution of zeros for systems comprising only 30 to 55 particles.
The question about further phase transitions to states with several
vortices remains open.

\section{Acknowledgements}
We acknowledge discussions with G. M. Kavoulakis, B. Mottelson, and
S. M. Reimann.  TP acknowledges research support as a Eugene P. Wigner
Fellow and staff member at the Oak Ridge National Laboratory (ORNL); this
research used resources of the Center for Computational Sciences at
ORNL. ORNL is managed by UT-Battelle, LLC for
the U.S. Department of Energy under Contract DE-AC05-00OR22725.

\clearpage
\onecolumn

\begin{table}[b]
\begin{tabular}{cccc}
$N$ & $\tau_1$ & $\gamma$ & $\alpha$ \\
\hline
40 & 0.0715 & -0.03  & -0.1109 \\
50 & 0.055  &  0.0   & -0.145  \\
52 & 0.053  &  0.003 & -0.147  \\
55 & 0.050  &  0.003 & -0.148  \\
\end{tabular}
\caption{ Calculated parameters $\tau_1$, $\gamma$,
and $\alpha$ for the various boson systems at $L/N=1$.}
\label{table:params}
\end{table}

\end{document}